\documentclass[runningheads]{svmult}

\usepackage{makeidx}   
\usepackage{graphicx}  
\usepackage{subeqnar}  
\usepackage{multicol}  
\usepackage{physmubb}  
\makeindex             

\newcommand{\gev}{{\rm Ge}\kern-1.pt{\rm V}}

\newcommand{\sla}[1]{/\!\!\!#1}

\newcommand{\menor} {\mbox{\raisebox{-0.4ex}
{$\;\stackrel{<}{\scriptstyle \sim}\;$}}}

\def\NPB{{\em Nucl. Phys.} B}
\def\PLB{{\em Phys. Lett.}  B}

\def\PRD{{\em Phys. Rev.} D}

\begin{document}
\title*{Prospects of Higgs Physics at the LHC \thanks{Talk given on behalf of the CMS and ATLAS collaborations at the \emph{14th Topical Conference on Hadron Collider
       Physics} (HCP2002), Karlsruhe, Germany, 30 Sep--4 Oct 2002. } }

\titlerunning{Prospects of Higgs Physics at the LHC}

%
%
%
%
%

\author{Bruce Mellado \\
\textit{University of Wisconsin - Madison 
\\ Department of Physics } }

\authorrunning{Bruce Mellado}

%
%
%

\maketitle              

\vspace{-1.cm}
\abstract{The search for the  Higgs boson is a major physics goal of the future Large Hadron Collider. The discovery potential is described as a function of the Higgs mass. It is shown that a Standard Model Higgs boson can be discovered within the first year of data taking. The status of MSSM Higgs searches is also discussed.}

\section{Introduction}

A major physics priority of the experiments at the future proton-proton Large
Hadron Collider (LHC) at CERN will be the search for the
Higgs boson, a cornerstone for the study of the nature of electroweak symmetry breaking.
Two general purpose experiments under construction, CMS~\cite{cms} and ATLAS~\cite{atlas},  have been optimized to
cover a large spectrum of possible physics signatures within the data taking environment of  the LHC.

Detailed simulations of both experiments have been performed to
demonstrate  the feasibility of the discovery  of the Standard Model (SM) Higgs in a broad range of the SM Higgs masses. The LHC experiments have also a large potential to explore the  Minimal Supersymmetric Standard Model (MSSM) Higgs sector.

\vspace{-0.2cm}
\section{Running Conditions and Physics Analysis}

Physics studies in CMS and ATLAS  assume an initial instantaneous luminosity of
$10^{33} \mbox{cm}^{-2} \mbox{s}^{-1}$ (low luminosity regime) for the turn on of the LHC. The instantaneous luminosity is expected to increase gradually before reaching the designed value of 
$10^{34} \mbox{cm}^{-2} \mbox{s}^{-1}$ (high luminosity regime). A total integrated luminosity of 10~fb$^{-1}$ should be delivered during the first year of data taking. After the following year 30~fb$^{-1}$ will be delivered. Once the design luminosity is reached 100~fb$^{-1}$ will be available per year.

\begin{figure}
\begin{center}
\hspace{-0.2cm}
\includegraphics[width=.5\textwidth]{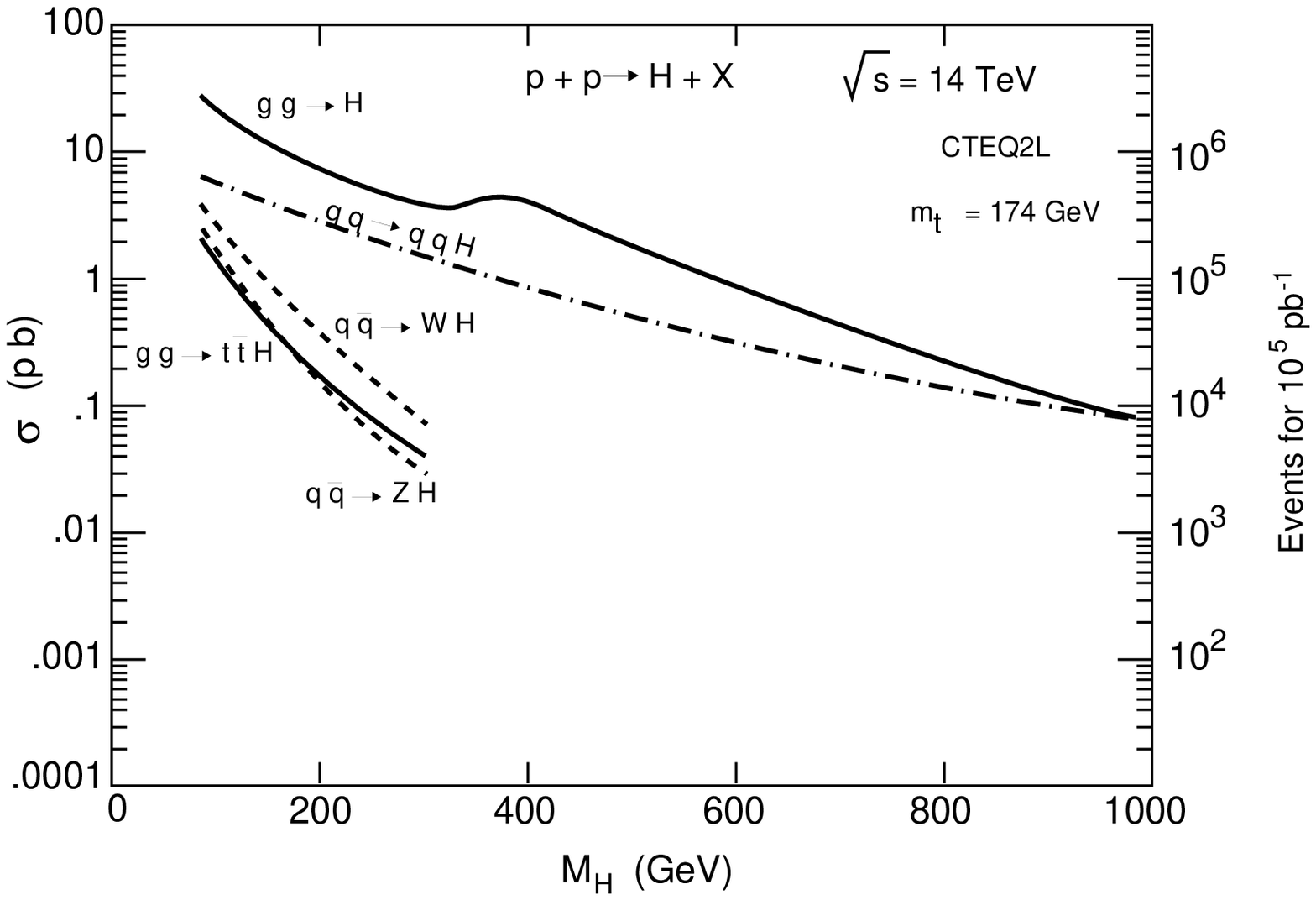}
\includegraphics[width=.45\textwidth]{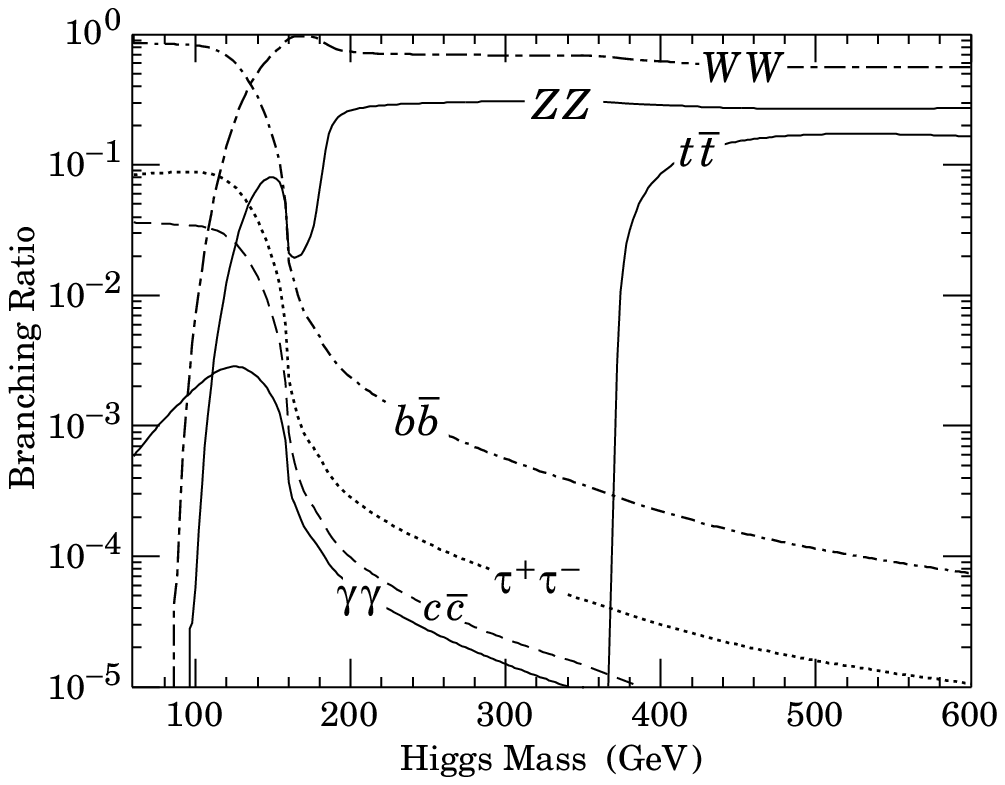}
\end{center}
\vspace{-0.6cm}
\caption[]{Higgs production cross section at the LHC  and Higgs
branching ratios as a function of the Higgs mass, $M_H$.}
\label{fig:higgscross}
\end{figure}

A number of elements are specific to the physics performance studies reported here:
\begin{itemize}
\vspace{-0.13cm}
\item The non diffractive $p-p$ inelastic cross section is assumed to be 70~mb. It is expected that an average of $\approx 2$ and $\approx 20$ inelastic collisions will occur together with the hard interaction for low and high luminosity regimes, respectively~\footnote{This effect is usually referred to as pile-up.}. This effect are included for both the low and high luminosity regimes.
\item Born-level cross-sections for both signal and backgrounds are used. This is motivated by the fact that the higher order QCD corrections (K-factors) to the  Born-level cross-sections of a number of meaningful processes are still unknown.
\item Parton level MC generators are interfaced with the PYTHIA package~\cite{pythia} for the simulation of partonic showers, hadronization and particle decays.
\item In many studies fast simulations of the response of the detectors is used. Nevertheless, the fast simulations are based on a full GEANT simulation of the detectors.
\end{itemize}
\vspace{-0.13cm}

\vspace{-0.2cm}
\section{The search for the SM Higgs boson}

The SM Higgs boson will be produced at the LHC predominantly via gluon-gluon fusion. This is illustrated in the left plot of  Fig.~\ref{fig:higgscross}, in which the expected Born-level cross-sections for each individual mechanisms are plotted as a function of the Higgs mass, $M_H$~\cite{spira}. For $M_H>100\,\gev$ the second dominant production process goes via the vector boson fusion (VBF). The associated production of SM Higgs with other heavy particles such as $W$, $Z$ or $t\overline{t}$ pair is sizable for $M_H \leq 200\,\gev$. The right plot of Fig.~\ref{fig:higgscross} displays the SM Higgs branching ratios as a function of $M_H$~\cite{spira}.

\begin{figure}
\begin{center}
\hspace{-0.2cm}
\includegraphics[width=.43\textwidth]{sens30.epsi}
\includegraphics[width=.484\textwidth]{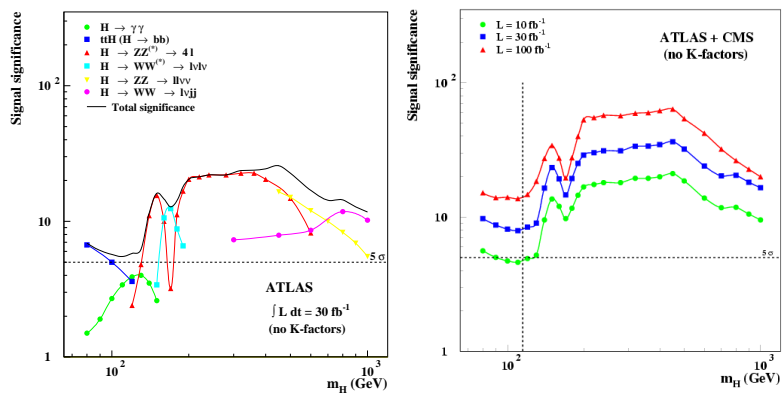}
\end{center}
\vspace{-0.7cm}
\caption{
Sensitivity for the discovery of a SM Higgs boson as a
function of $M_H$. The statistical significances are
plotted for individual channels, assuming an integrated luminosity of
30~fb$^{-1}$ (left) for the ATLAS experiment. The overall statistical
significance is plotted (right) as a function of the Higgs mass for  10, 30, and 100~fb$^{-1}$ for
CMS and ATLAS combined.}
\label{fig:sensitivity}
\end{figure}

Most of the relevant discovery modes  are well assessed~\cite{cms,atlas}.
The overall sensitivity for the discovery of a SM Higgs boson
over the a large mass range $80<M_H<1000\,\gev$ is shown in the left plot of 
Fig.~\ref{fig:sensitivity} for one single experiment (ATLAS) and 30~fb$^{-1}$ of accumulated luminosity. The combined sensitivity for both CMS and ATLAS is shown in the right plot of 
Fig.~\ref{fig:sensitivity} for 10~fb$^{-1}$, 30~fb$^{-1}$ and
100~fb$^{-1}$. The range of $M_H$ close to the LEP limit~\cite{LHCC01-55} remains a challenging one.

The strategy adopted for the SM Higgs searches depends upon $M_H$~\cite{cms,atlas}:
\begin{itemize}
\vspace{-0.13cm}
\item The low mass region, $M_H<130\,\gev$. 
Excellent  energy resolution and
background rejection are required. Two decays modes are
important in this $M_H$ region: $H \rightarrow b\overline{b}$ and $H
\rightarrow \gamma \gamma$. The branching ratio of $H \rightarrow b\overline{b}$ is close to
1. However, due to the overwhelming QCD background the signal-to-background ratio for the inclusive
production is smaller than $10^{-5}$.
The $H \rightarrow b\overline{b}$ decay mode may be used in the associated production of Higgs with $W$, $Z$, or $t\overline{t}$ pair by requiring in the final state a charged lepton ($e,\mu$).
The $H \rightarrow \gamma \gamma$ channel has a small branching ratio ($\propto 10^{-3}$) but the expected signal-to-background ratio ($\approx 10^{-2}$) makes this channel interesting for inclusive searches.

 \item The intermediate mass region, $130\,\gev<M_H<2 M_Z$. 
The most powerful channels are $H \rightarrow ZZ^{(*)} \rightarrow
4l$  and $H \rightarrow WW^{(*)} \rightarrow l \nu l \nu$. The latter can be used in 
inclusive and associated W production.
\item The High mass region, $M_H >M_Z$. Here the discovery is most straightforward thanks to 
the $H \rightarrow ZZ \rightarrow 4l$ channels. In this case, a narrow resonance is expected in the presence of little background. For very heavy masses ($M_H>500\,\gev$) the production cross-section decreases noticeably. To compensate this the decay modes $H \rightarrow ZZ \rightarrow ll \nu \nu$
and $H \rightarrow WW \rightarrow l \nu~2 jets$ are used.
\end{itemize}
\vspace{-0.13cm}

\begin{figure}
\begin{center}
\hspace{-0.2cm}
\includegraphics[width=.5\textwidth]{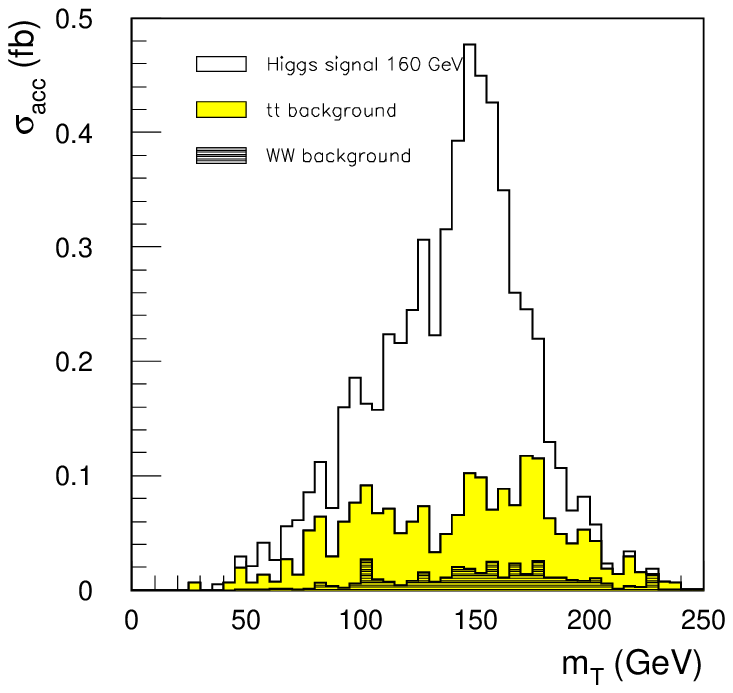}
\includegraphics[width=.5\textwidth]{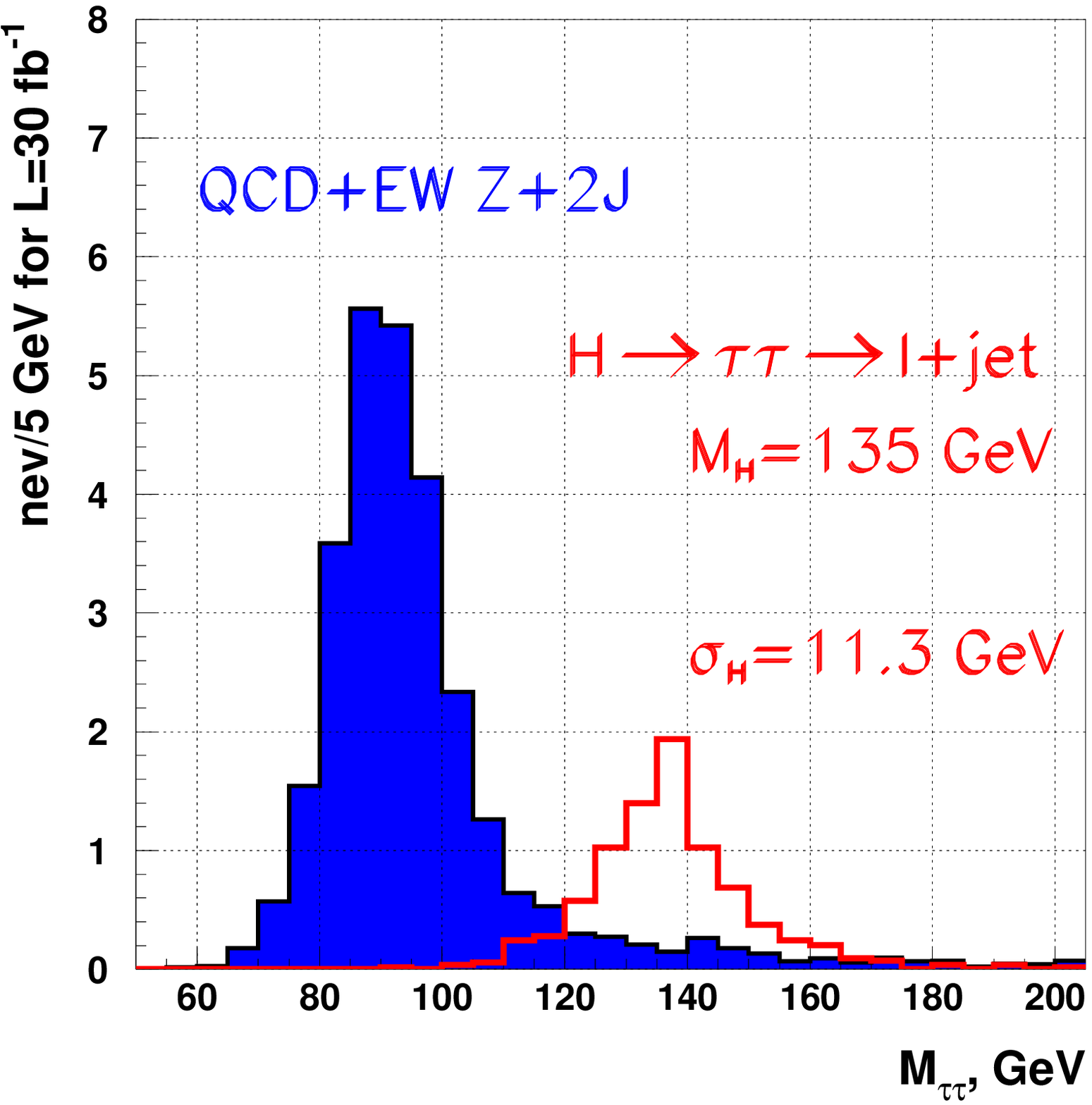}
\end{center}
\vspace{-1.1cm}
\caption[]{The left plot shows the expected $M_T$ distribution for  $M_H=160\,\gev$ with $H\rightarrow WW^{(*)} \rightarrow l^{+}l^{-}\sla{p_{T}}$. The accepted cross-sections $d\sigma/dM_T$ (in fb/$5\,\gev$)  are shown. The yellow and black histograms correspond to the contribution from the $t\overline{t}$ and $WW$ backgrounds, respectively. The right plot shows the distribution of the invariant mass of the $tau$-pair for $M_H=135\,\gev$ with $H\rightarrow \tau\tau \rightarrow l+jet$ (in events/$5\,\gev$) for an accumulated luminosity of 30 fb$^{-1}$. The blue and red histograms correspond to the contribution from the major backgrounds (see text) and the signal, respectively.
}
\label{fig:vbfmass}
\end{figure}

\vspace{-0.2cm}
\subsection{Recent Progress in SM Higgs Searches}

In earlier physics performance studies the searches for the SM Higgs with two forward jets~\footnote{This enhances the relative contribution from the VBF mechanism.} was considered for $M_H>300\,\gev$~\cite{atlas}. The bulk of the recent progress in performance studies for SM Higgs searches is related to the inclusion of VBF modes in the low $M_H$ region~\cite{LesHouches2002,CMS-2002/016,ATL-PHYS-2002-018,NIKITENKO,ATLASSN}.

Analyses performed at the parton level with  $H\rightarrow WW^{*}$ and $H\rightarrow \tau\tau$ suggested that the VBF modes could be the most powerful discovery modes in the range $115<M_H<200\,\gev$~\cite{pr_60_113004,pr_61_093005,pl_503_113}.
The analyses for the $WW^{*}$ and $\tau\tau$ Higgs decay modes have been redone. A more detailed simulation of the  LHC detectors has been implemented. This includes the forward jet tagging and the central jet veto efficiencies. The physics performance had been assessed for the low luminosity regime and for integrated luminosity values of up to 30~fb$^{-1}$.

The basic signature of this process with $WW^{*}$ and $\tau\tau$ Higgs decay modes is: 
\begin{itemize}
\vspace{-0.13cm}
\item Two energetic jets in the forward detectors (tagging jets).
\item Suppressed hadronic activity between tagging jets.
\end{itemize}
\vspace{-0.13cm}

Because of trigger rate requirements decay modes are chosen with one or two large transverse momentum electrons or muons. The following background processes common to all modes have been considered:
\begin{itemize}
\vspace{-0.13cm}
\item Production of $t\overline{t}$: At leading order this process contributes as a background due to the two b-jets. Parton emission plays an important role here.
\item QCD WW production: The continuum production of W pairs associated with two jets identified as tagging jets.
\item Electroweak (EW) WW production: Production of W pairs via de t-channel vector boson exchange associate with jets. The cross-section of this process is small but rejection is harder due to similarities of the final state topologies with the signal processes.
\item QCD Drell-Yan associated with jets with $Z/\gamma^{\star}\rightarrow l^+l^-$.
\item EW $\tau\tau$ production associated with jets. 
\end{itemize}
\vspace{-0.13cm}

\begin{figure}
\begin{center}
\hspace{-0.2cm}
\includegraphics[width=.5\textwidth]{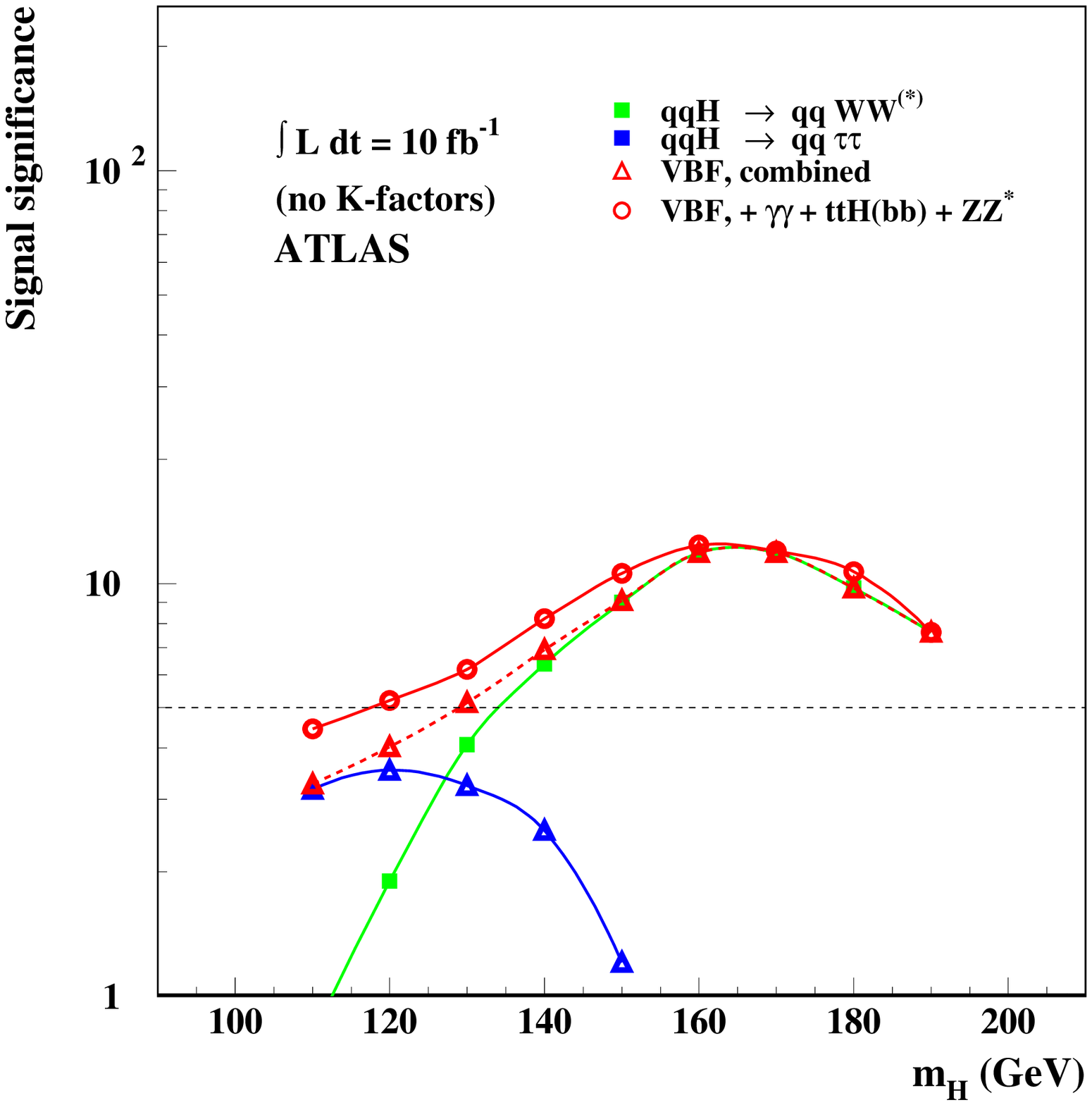}
\includegraphics[width=.5\textwidth]{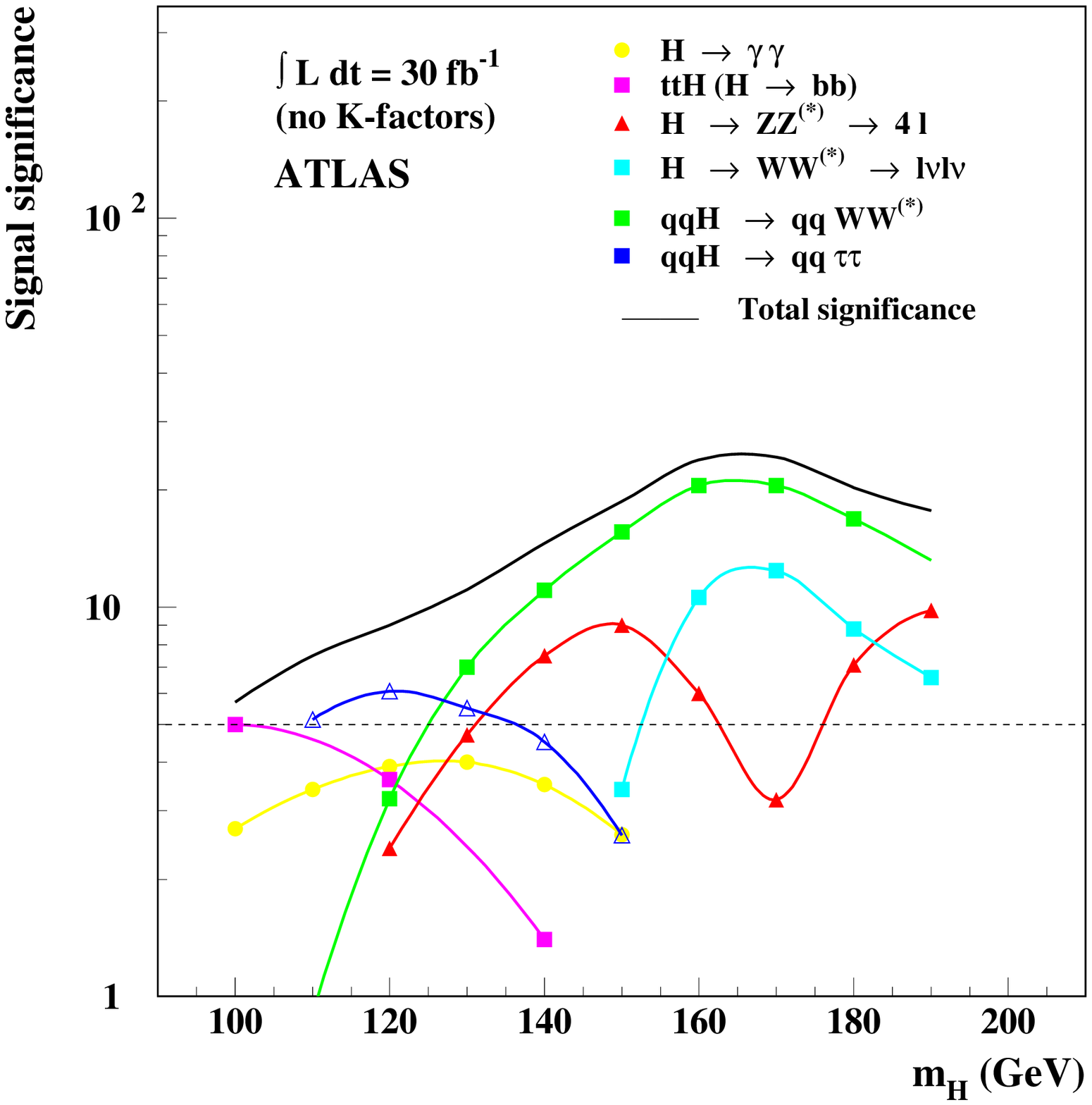}
\end{center}
\vspace{-0.8cm}
\caption[]{Impact at low $M_H$ of VBF channels on the sensitivity of a single experiment. The left and right plots show the signal significance for one experiment alone (ATLAS) for 10~fb$^{-1}$ and 30~fb$^{-1}$ of accumulated luminosity, respectively. }
\label{fig:vbfsignif}
\end{figure}

The EW WW and  $\tau\tau$ backgrounds have been generated with dedicated programs~\cite{MadCUP}. The Drell-Yan background has been generated using the matrix element calculations for $q\rightarrow Zg$ and $qg\rightarrow Zg$ provided by the PYTHIA package. 

In the $H\rightarrow WW^{*}$ analysis a large rejection  may be achieved against the $t\overline{t}$ and the WW backgrounds by taking advantage of the anti-correlation of the spins of the W's of the decay of the Higgs~\cite{pr_55_167}. In this case no mass peak is seen. Instead an excess of events in the transverse mass, $M_T$, is observed~\cite{pr_60_113004,pl_503_113}. The left plot in Fig.~\ref{fig:vbfmass} displays the expected $M_T$ distribution for a Higgs of $M_H=160\,\gev$ with $H\rightarrow WW^{(*)} \rightarrow l^{+}l^{-}\sla{p_{T}}$. The accepted cross-sections $d\sigma/dM_T$ (in fb/$5\,\gev$) including all efficiency and acceptance factors are shown. The yellow and black histograms correspond to the contribution from the $t\overline{t}$ and $WW$ backgrounds, respectively.
In the $H\rightarrow \tau\tau$ analysis the invariant mass of the $\tau\tau$, $M_{\tau\tau}$,  is reconstructed using the collinear approximation~\cite{np_297_221}. A resolution of $\approx 10\,\gev$ is achieved for low mass Higgs. The right plot in Fig.~\ref{fig:vbfmass} shows the distribution of $M_{\tau\tau}$ for $M_H=135\,\gev$ with $H\rightarrow \tau\tau \rightarrow l+jet$. The number of expected events in bins of $5\,\gev$ for an accumulated luminosity of 30 fb$^{-1}$ is shown. The blue and red histograms correspond to the contribution from the QCD and EW $Z$ associated with two or more jets and the signal, respectively.

Fig.~\ref{fig:vbfsignif} illustrates the impact of VBF channels on the sensitivity of a single experiment to low $M_H$. The left and right plots show the signal significance for one experiment alone (ATLAS) for 10~fb$^{-1}$ and 30~fb$^{-1}$ of accumulated luminosity, respectively. A $5\sigma$ signal significance may be reached for a single experiment with VBF channels combined for $130<M_H<190\,\gev$. When combined with other relevant modes a single experiment can reach a signal significance well above $5\sigma$ in the entire range of the Higgs search for 30~fb$^{-1}$ of accumulated luminosity.

\vspace{-0.2cm}
\section{Recent Progress in MSSM Higgs Searches}

\begin{figure}
\begin{center}
\hspace{-0.2cm}
\includegraphics[width=.42\textwidth]{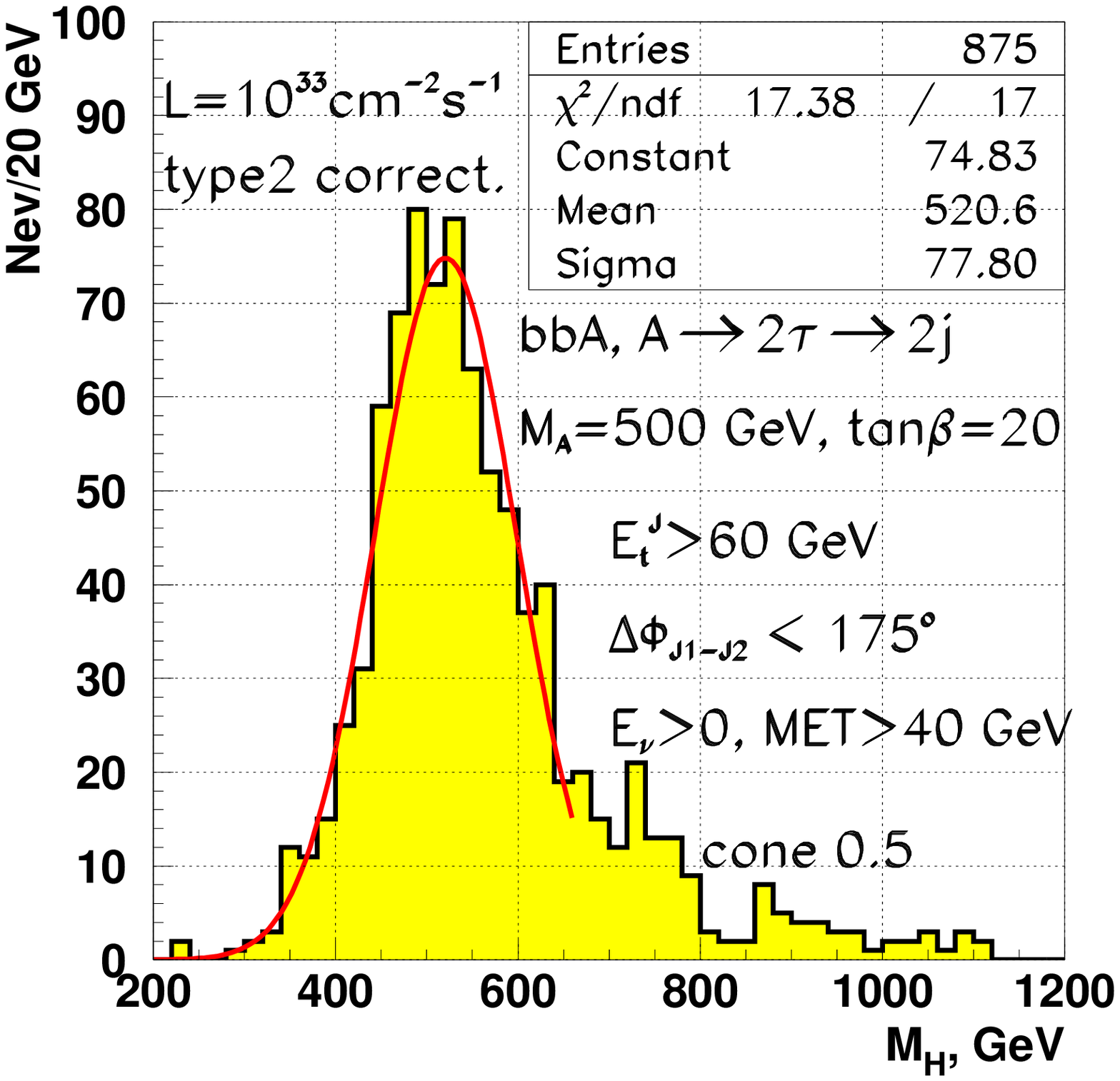}
\includegraphics[width=.57\textwidth]{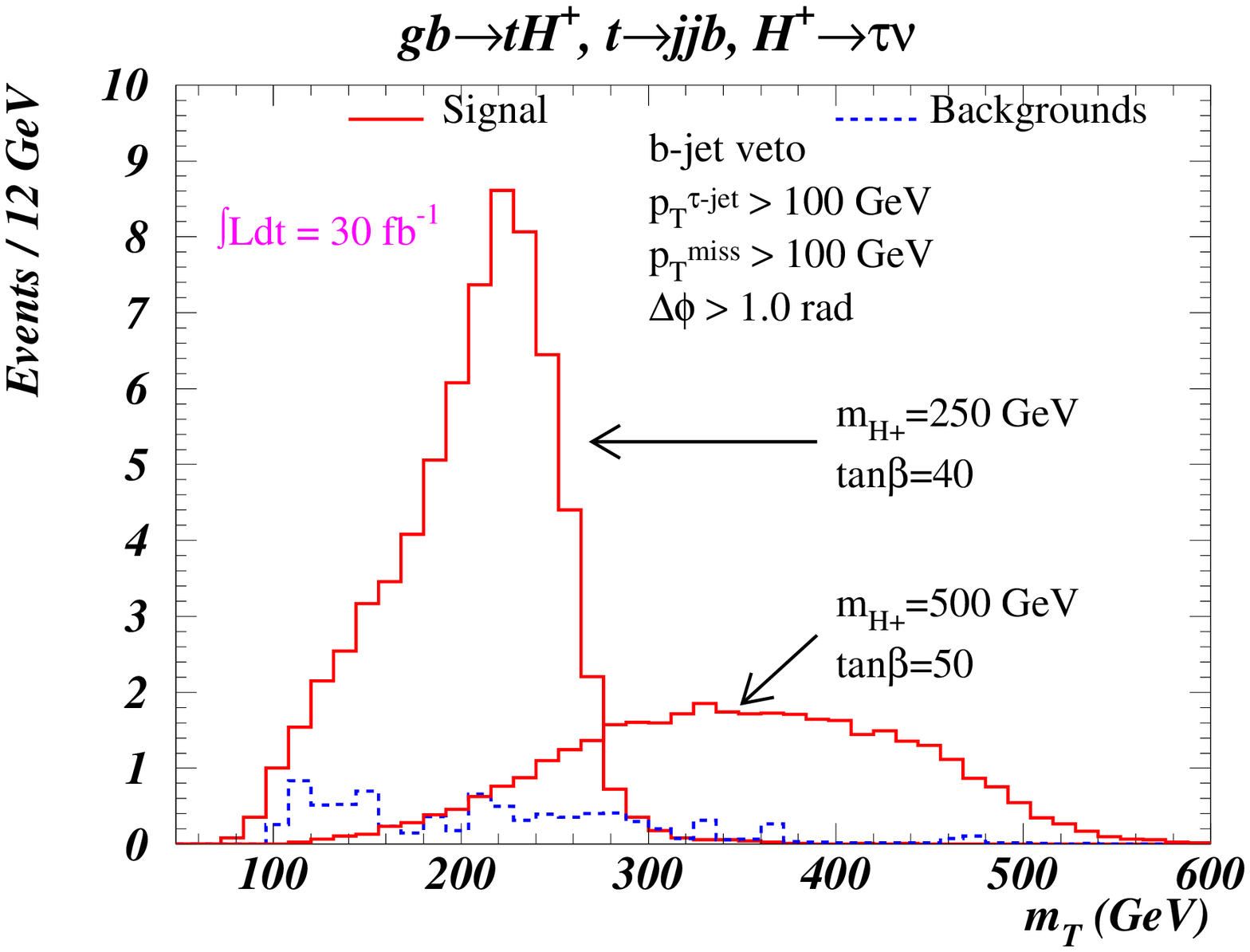}
\end{center}
\vspace{-0.8cm}
\caption[]{The left plot shows the reconstructed invariant mass of the $\tau\tau$ (events/20$\,\gev$)  for $b\overline{b}, H\rightarrow \tau\tau \rightarrow 2\tau $jets for $M_H=500\,\gev$  and $\tan{\beta}=20$. The plot on the right displays the expected $M_T$ (events/12$\,\gev$) with $gb\rightarrow tH^+, t\rightarrow jjb, H^+\rightarrow\tau\nu$ for $M_{H^+}=250, 500\,\gev$. The dashed line corresponds to the background contribution. Both plots correspond to 30~fb$^{-1}$ of accumulated luminosity. 
}
\label{fig:mssmhighmass}
\end{figure}

With standard assumptions the $M_{A}/\tan{\beta}$ plane should be covered by the LHC experiments with the possibility of observing more than one MSSM Higgs in a large portion of the plane~\cite{cms,atlas}. Significant progress has been reported on the study of modes specific to the MSSM Higgs sector, namely, $A/H\rightarrow \tau\tau$  with two $\tau-$jets in the final state~\cite{LesHouches2002}. The left plot in Fig.~\ref{fig:mssmhighmass} shows the reconstructed invariant mass of the $\tau\tau$ for $b\overline{b}, H\rightarrow \tau\tau \rightarrow 2\tau-$jets for $M_H=500\,\gev$ and $\tan{\beta}=20$ with full detector simulation. The resolution of the reconstructed Higgs is $\approx 15\%$ in the mass range under study. In the case of charged Higgs no mass peak is reconstructed. Instead an excess of events in the $M_T$ distribution is expected. The plot on the right  in Fig.~\ref{fig:mssmhighmass} displays the reconstructed $M_T$ with $gb\rightarrow tH^+, t\rightarrow jjb, H^+\rightarrow\tau\nu$ for $M_{H^+}=250, 500\,\gev$. The contribution from SM backgrounds may be heavily suppressed by taking advantage of the polarization of the decaying $\tau$'s.

\begin{figure}
\begin{center}
\hspace{-0.2cm}
\includegraphics[width=.5\textwidth]{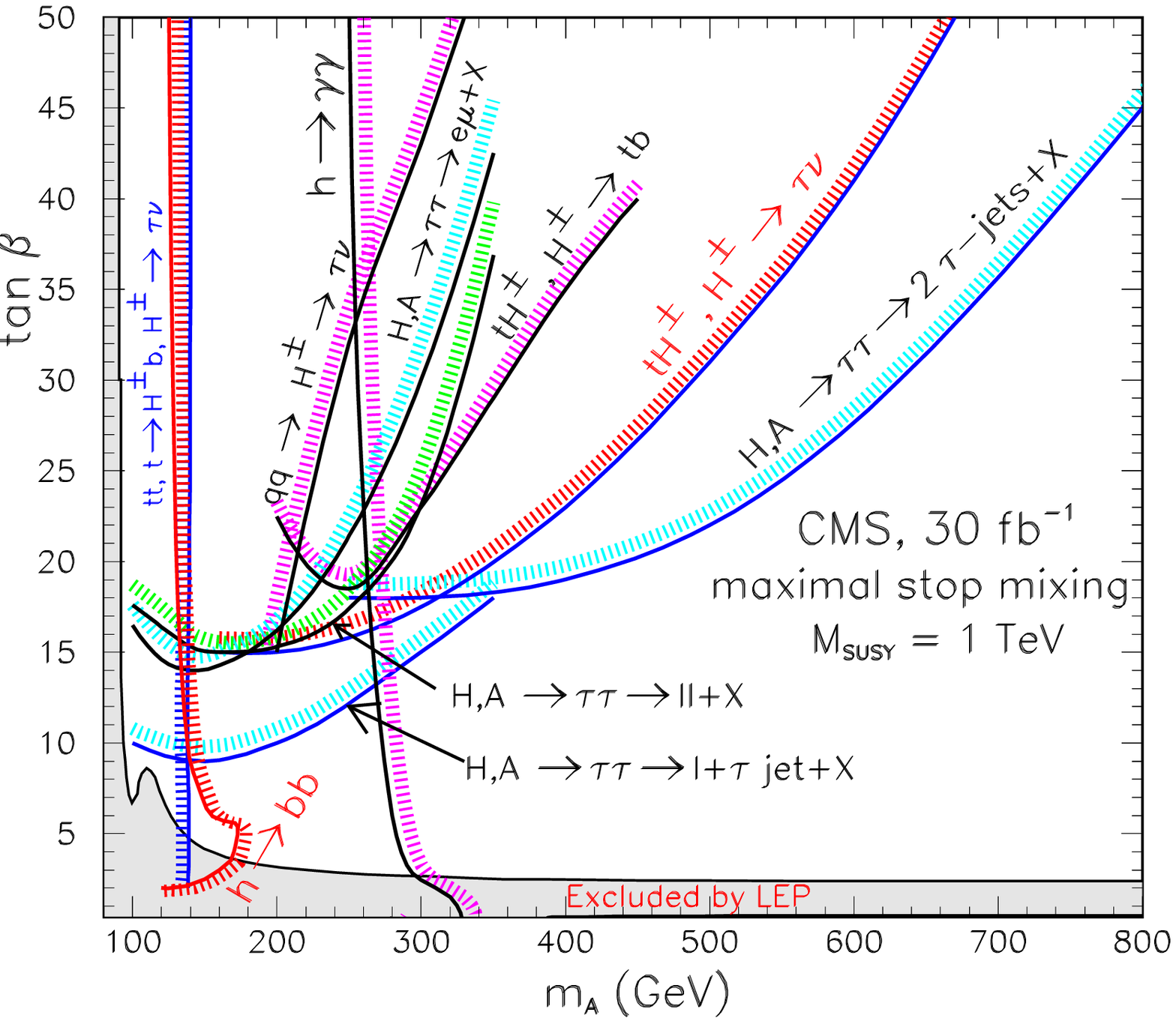}
\includegraphics[width=.45\textwidth]{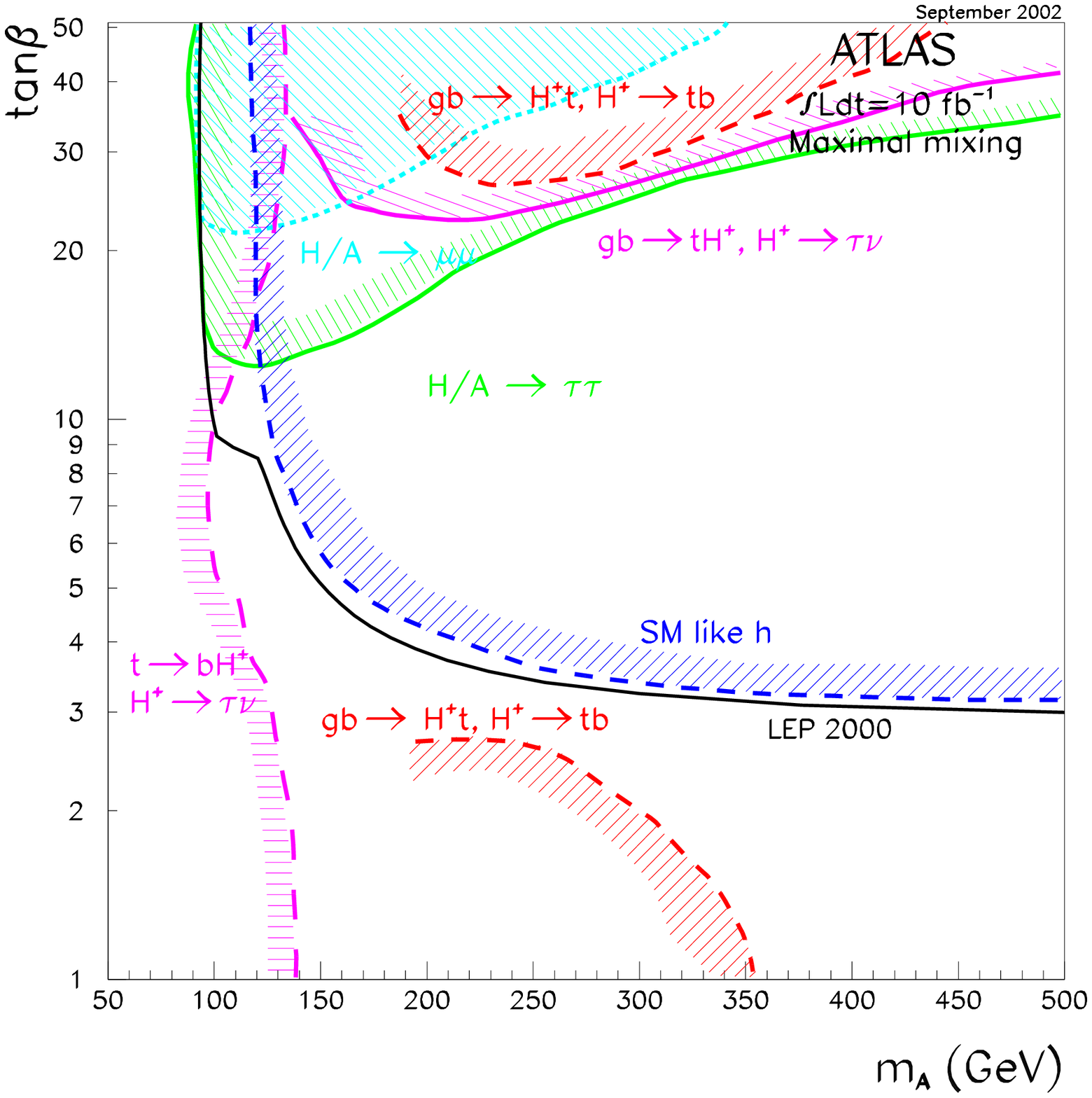}
\end{center}
\vspace{-0.6cm}
\caption[]{The plot on the left shows the expected the 5$\sigma$ contours in the $M_{A}/\tan{\beta}$ plane for the MSSM Higgs with CMS for 30~fb$^{-1}$ of accumulated luminosity. The shaded area is excluded by LEP~\cite{LEPMSSM}. On the right the 5$\sigma$ contours for CMS and ATLAS combined for 10~fb$^{-1}$ of accumulated luminosity are shown (see text).}
\label{fig:mssmsignif}
\end{figure}

The plot on the left in Fig.~\ref{fig:mssmsignif} shows the expected 5$\sigma$ contours in the $M_{A}/\tan{\beta}$ plane for the MSSM Higgs for one experiment (CMS) in the minimal mixing scenario for 30~fb$^{-1}$ of accumulated luminosity. The improvement in the coverage is mainly due to the inclusion of modes with two $\tau-$jets in the final state. The plot on the right in Fig.~\ref{fig:mssmsignif} corresponds to the 5$\sigma$ contours for CMS and ATLAS combined for 10~fb$^{-1}$ of accumulated luminosity. The line marked with {\em SM like h} includes the combination of the SM model like channels, including VBF. This illustrates that most of the MSSM plane will be covered by the two LHC experiments with just 10~fb$^{-1}$ of accumulated luminosity. 

Because the suppression of the MSSM Higgs coupling to weak bosons the region of low and medium $\tan\beta$ remains a difficult one for heavy Higgs. Scenarios  in which the scale of supersymmetry (SUSY) is low enough so as to allow  Higgs decay into SUSY particles or Higgs production from SUSY particle cascades are under investigation. This enhances the sensitivity of the LHC experiments to the low and medium $\tan\beta$ region.

\vspace{-0.2cm}
\section{Conclusions}

The status of performance studies for SM and MSSM Higgs bosons searches at the LHC is discussed. The inclusion of VBF modes in the search for SM Higgs improves the expected signal significance for low and medium $M_H$. One single experiment may achieve a $5\sigma$ signal significance for the range $115<M_H\menor1000\,\gev$ with just 10~fb$^{-1}$ of accumulated luminosity.

The coverage of the MSSM plane with more than one Higgs  has been significantly enhanced by including heavy Higgs decays into $\tau$'s with $\tau-$jets in the final state. Most of the MSSM plane will be covered by the two LHC experiments with just 10~fb$^{-1}$ of accumulated luminosity. Studies have been performed to determine the impact of the inclusion of Higgs decays into SUSY particles and Higgs from SUSY cascades to cover the intermediate and low $\tan{\beta}$ range.

\end{document}